# Computational Investigations into the Origins of Short-Term Biochemical Memory in T cell Activation

Jason W. Locasale*

Department of Biological Engineering, Massachusetts Institute of Technology, Cambridge, Massachusetts, United State of America

Recent studies have reported that T cells can integrate signals between interrupted encounters with Antigen Presenting Cells (APCs) in such a way that the process of signal integration exhibits a form of memory. Here, we carry out a computational study using a simple mathematical model of T cell activation to investigate the ramifications of interrupted T cell-APC contacts on signal integration. We consider several mechanisms of how signal integration at these time scales may be achieved and conclude that feedback control of immediate early gene products (IEGs) appears to be a highly plausible mechanism that allows for effective signal integration and cytokine production from multiple exposures to APCs. Analysis of these computer simulations provides an experimental roadmap involving several testable predictions.

## INTRODUCTION

The orchestration of the adaptive immune response is predicated on the integration of signals derived from peptide fragments that bear the molecular signature of an invading pathogen. T cells become activated by integrating signals derived from such peptides that are presented with proteins from the Major Histocompatibility complex (MHC) that are present on the surface of antigen presenting cells (APCs). Upon engagement with an APC, signal transduction is initiated by the interaction of the T cell Receptor (TCR) with MHC-peptide complexes. Successful signal integration results in numerous phenotypic outcomes and allows for the T cell to coordinate an appropriate immune response. In contrast, the failure of such priming processes leads to deleterious consequences such as autoimmunity.

The advent of two-photon imaging technologies has allowed for the study of real-time, in vivo T cell activation in lymph nodes in the presence of an antigenic challenge[1]. One consequence of such technological advancements is that recent imaging experiments have challenged the idea that the activation of naïve T cells requires prolonged continuous exposure from a single APC. These experiments suggest that, under certain conditions, T cells may integrate signals from short interrupted exposures to antigen presentation. For instance, in vivo mouse studies show that, during the activation process, there exists a stage where T-cells are involved in multiple transient interactions with many APCs[2].

In a parallel investigation into the nature of T cell signaling, a Control T cell Activation dithFseries of provocative *in vitro* experiments were conducted to specifically address the question of whether T-cells can integrate multiple interrupted signals and utilize the accumulation of these signals for a biological response[3]. In their model system, signaling between Th1 T-cells and B-cell APCs was initiated in a collagen matrix. IFN-$\gamma$ production along with other T-cell signaling markers such as calcium mobilization, ERK activity, and immunological synapse formation were monitored. Conjugation with APCs resulted in a sharp rise in calcium mobilization and ERK activity. In their system, IFN-$\gamma$ production commences after roughly 30 minutes of active signaling through the cell-cell contact. The authors assessed whether T cells can integrate interrupted signals by introducing a reversible src family kinase inhibitor, PP2, after signaling had begun upon T cell–APC conjugation. This inhibitor is known to have a high selectivity towards Lck, a crucial src-family kinase responsible for triggering downstream pathways in T cells.

Introduction of PP2 quickly abrogated both calcium mobilization and ERK activity, and the immunological synapse also rapidly disassembled. After a prescribed time interval of approximately 20 minutes, PP2 was washed out of the culture and signaling, as determined by calcium flux and ERK activation, resumed almost immediately; the immunological synapse also reassembles.

In addition, the authors report several other unexpected findings. They first demonstrate that thirty minutes of stimulation is initially insufficient for cytokine (IFN-$\gamma$, in this case) production. They then show that subsequent rounds of TCR signaling, after interruption of the signal, produce significant amounts of cytokine within thirty minutes. These results suggest that T-cells have the ability to integrate interrupted signals from multiple encounters with antigen and also suggests that T-cells can exhibit a "short-term" (the phrase "short-term" is used to distinguish from the 'long-term' development of the T cell memory phenotype) memory of past exposures to antigen in that the first exposure seems to prepare the T cell for subsequent exposures to antigen. Fig. 1 gives a schematic of the main findings. Such a memory in this case persists for at least 20 minutes and moreover, there is also evidence to suggest, in other situations, that such a memory can persist for much longer periods of time[4]. The mechanistic origin of such a short-term memory is not understood although it is likely the result of the sustained activation of some signaling intermediate. Such a sustained activity can provide a means by which to integrate signals from transient cell-cell contacts and other types of interrupted signaling. Therefore, such an understanding may have direct applications related to the nature of

.....................................................................

Funding: Work was undertaken through the sponsorship of Professor Arup Chakraborty who was able to support the work with funds provided through an NIH director's pioneer award

Competing Interests: The authors have declared that no competing interests exist.

* To whom correspondence should be addressed. E-mail: Locasale@MIT.edu

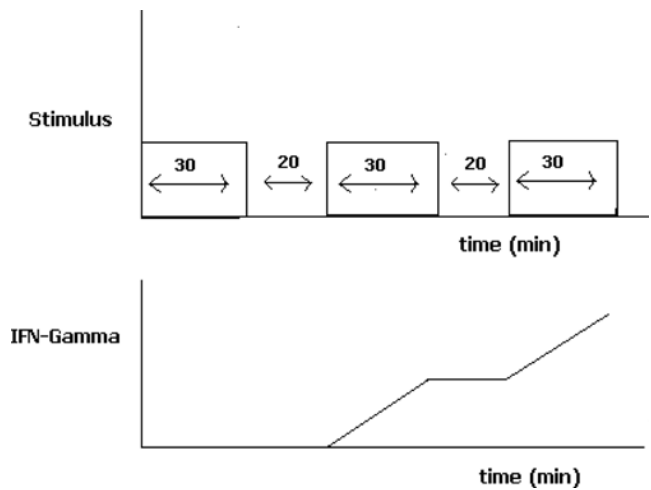

**Figure 1. Schematic of experimental results obtained demonstrating memory in signal integration.** a.) TCR mediated signal transduction proceeds for thirty minutes via a contact with APCs in a collagen matrix. After thirty minutes, signaling is aborted with the T-cell-APC contact intact. Following a period of approximately 20 mins, signaling is allowed to resume. The procedure is repeated periodically. b.) Cytokine (i.e. IFN-$\gamma$) production is measured. The first 30 minutes of signaling results in no IFN-$\gamma$ production. Upon the next 30 minute round of TCR signaling, IFN-$\gamma$ is produced. The final round of stimulation results in IFN-$\gamma$ production as well. Curiously, the first round of signaling, while insufficient for IFN-$\gamma$ production, prepares (i.e. establishes a memory) for more rapid IFN-$\gamma$ production in subsequent rounds of stimulation.
doi:10.1371/journal.pone.0000627.g001

transient versus stable T cell-APC contacts during T cell priming and activation.

The acquisition of 'short-term' memory is an emergent property that requires the coordination of cooperative, dynamic events involving many interacting cellular components. Thus, it is often difficult to intuitively assess how different molecules act in concert to exhibit collective, system-wide cellular behaviors such as signaling memory. In this regard, theoretical and computational studies[5] that involve mathematical modeling and allow for the systematic parsing of many dynamical processes have proven useful as a supplement to genetic, biochemical, and imaging experiments.

In the scope of this work, we present results from computer simulations of mathematical models that explore the physiological consequences of different mechanisms that can account for the biochemical memory observed in several experiments. We first briefly discuss some key observations from existing data in the literature and in particular, the experiment[3] that allowed us to construct our mathematical models. Next we present results from our modeling efforts that involve stochastic computer simulations of simplified versions of the signaling pathways leading to cytokine production.

Our goal was not to simulate every detail of the T-cell signaling network but rather to construct the simplest possible models that can investigate how the basic structure of several different molecular mechanisms could lead to the collective property of biochemical memory in T cell signaling. Other computational models have considered many elements of T cell activation such as how sensitivity to antigen is established[6,7]. For example, 'digital' Erk responses that arise from the coupling of positive and negative feedback loops has been extensively studied as a model for signal integration at short time scales[6,8]

Instead, our modeling efforts specifically aim to demonstrate how the first round of signaling can seemingly prepare the T cell for signal integration so that later exposures to antigen do not require a long lag time for cytokine production. Towards this end, we focus on events at later times that occur downstream of Erk activation. We derive from these models, testable predictions that can ascertain many aspects of such memory phenomena in T cell signal integration. The predictions that emerge from our calculations suggest several experiments that could further elucidate the possible mechanism for this 'short-term' biochemical memory. Finally, it is our hope that such calculations serve as a template for further quantification and modeling of memory phenomena and signal integration that are observed in T cell signaling.

### Possible sources for 'short-term' signaling memory

While initial signaling events, such as the mobilization of intracellular calcium and activation MAPK pathways, occur within minutes of the initial T cell/APC contact, at least thirty minutes of signaling is required for cytokine production. One hypothesis for the existence of this waiting period is that there is a time required for the accumulation of immediate early gene products (IEGs) such as Jun and Fos proteins which comprise the AP1 transcription factor complex. IEGs are synthesized de novo upon TCR signaling and their presence is a necessary condition for cytokine production[9]. T cells then must undergo signaling for long time periods on the order of hours in order to become fully activated[10].

Studies on the duration of Erk signaling leading to the accumulation of IEGs suggest that a hyper-phosphorylated state of the IEG product Fos can remain active for long times[11,12]. Furthermore, IEG products such as the family of Jun proteins have been observed to be active for long periods of time after the removal of TCR signals[13]. Hence, the activity of these transcription factors during periods of interrupted signaling could explain why the first round of signaling seems to prepare the T cell for cytokine production during later exposures to antigen. In this picture, the initial thirty minutes of signaling serve to accumulate IEG products that remain available for long times after the stimulus is removed. Then, for subsequent encounters with antigen, the activity of IEG products allows for faster cytokine production since this rate limiting step of the pathway is then bypassed. However, since all cytokine production ceases once TCR signaling is aborted, active IEGs alone can not be a sufficient condition for cytokine production. For instance, transcription factors, such as NFAT and NF-$\kappa$b, derived from the activation of the calcium pathway and the PKC-theta pathway are other necessary conditions for cytokine production[14,15].

One general way to generate sustained activity of signaling intermediates is to exploit positive feedback in signaling networks[16]. In a positive feedback loop, the end-product of the signaling pathway might up-regulate an activator of some upstream component of the pathway. Thus, once this activator is turned on, removal of the stimulus would not necessarily result in termination of the signal. Due to the autocatalytic nature of the feedback loop, provided that there is enough initial stimulation, active signaling intermediates can be self-sustained even in the absence of stimuli. This phenomenon has been termed bistability, hysteresis, or effective irreversibility. There are other ways in which bistability could be generated; for example, disrupting the activity of an inhibitor of the pathway can also result in bistability.

## RESULTS

### Mathematical models for biochemical memory

The first model that we studied is derived primarily from experimental results relating signal duration with the stability of IEG products[11]. Initial signaling events such as the activation of the calcium pathway and MAPK cascades occur within minutes of TCR-MHC engagement[17]. However, cytokine production does not commence immediately. As previously stated, one explanation for this waiting period is the time required for the accumulation of IEGs such as Jun and Fos proteins which comprise the AP-1 transcription factor complex. IEGs are synthesized *de novo* upon TCR signaling and are necessary for cytokine production[9]. Studies have shown that the duration over which activated Erk is maintained can be sensed by cFos, a protein product of IEGs; Erk can phosphorylate the Ser 362 and Ser 374 sites in cFos[11,18]. This form of phosphorylated cFos is unstable; but, it is primed for additional phosphorylation by Erk. A DEF domain in cFos docks Erk, and primed cFos can be then phosphorylated at the Thr 325 and Thr 331 sites. This hyper-phosphorylated form of cFos apparently remains active for long times. One simple hypothesis could be that hyper-phosphorylated cFos is not subject to inactivation. The sustained activity of the hyper-phosphorylated form of cFos, and hence the transcription factor AP1, during periods of interrupted signaling then explains why the first round of signaling seems to prepare the T cell for cytokine production during later exposures to antigen. In this model, the initial thirty minutes of signaling serve to accumulate IEG products which remain stable for long times following the removal of the stimulus. The accumulation could occur in a graded or switch-like manner as some have argued[19]. Then, for subsequent encounters with antigen, the activity of the IEG products allows for faster cytokine production since this rate limiting step of the pathway is removed from the signaling network.

However, since all cytokine production appears to cease once TCR signaling is aborted, active IEGs alone cannot be a sufficient condition for IFN gamma production. For example, calcium mobilization another necessary condition for cytokine production[15]. The calcium pathway ultimately leads to the activation of the transcription factor, NFAT, which then translocates into the nucleus. It has been documented that upon disruption of signaling, NFAT activity will quickly decay due to the presence of GSK–GSK phosphorylates NFAT which then signals NFAT to egress from nucleus[15]. This may be the reason why no cytokine production is observed once the signal is disrupted. If the initial period of signaling is short, then the stable form of cFos will not accumulate, and so T cells will be unable to add up signals from subsequent exposures to antigen. On the face of it, the observations discussed above support the speculation that the crux of the mechanism underlying the ability of T cells to integrate multiple interrupted signals is the creation of a hyper-phosphorylated stable form of cFos that is mediated by sustained Erk activation.

As it stands, this mechanism for signaling memory is not without difficulties. Degradation mechanisms that are mediated by various ubiquitin pathways have been shown to occur with members of the AP-1 complex such as Jun[20]. Moreover, turnover of IEG products mediated by ubiquitin pathways can be very fast in cells and is known in many cases to occur at rates faster than the duration of interrupted signaling that was measured. Furthermore, it is not clear why a hyperphosphorylated form of cFos cannot be dephosphorylated by phosphatases on a time scale much faster than the time during which the signal has been disrupted. Recognition of this potential difficulty leads to one hypothesis: IEG products such as Fos and Jun are embedded within positive feedback loops that allow their activity to persist long after the stimulus has been removed. Due to the autocatalytic nature of the feedback loop, an active signaling intermediate may be self sustained, even in the presence of protein degradation, by the catalytic cycle that is initiated within the signaling cascade. This hypothesis led us to investigate the biological consequences of models involving both the presence and absence of feedback loops.

## Computer simulations of the signaling models

The three scenarios examined in detail, are depicted in Fig. 2a. In each scenario active IEG product (e.g. cFOS) serves as the biochemical memory. Because the detailed biochemical mechanism by which cFOS is activated is not entirely known, we considered two cases. In the first case, Fig. 2b, the kinetics of cFOS phosphorylation are determined by laws of mass action involving a simple linear reaction mechanism. In the second case, Fig. 2c, the stabilization of cFOS by ERK is achieved cooperatively—the degree of cooperativity is determined by a Hill function. Lastly, in Fig. 2d, we consider the case where the hyperphosphorylated state of cFOS is maintained by positive feedback. A description of the network topologies used in the simulations as well as the kinetic parameters is given in the methods section and in Table 1. The sensitivity of the model to perturbations in the parameters used in the simulations is also discussed in the methods section.

The calculations aim to mimic the experiments by periodically interrupting signaling by "inhibiting Lck" in the simulation for a period and then removing the "inhibitor". This is accomplished by disallowing any contribution of triggered T cell receptors to the activation of downstream pathways for a specified time interval. The "strength" of the signal is determined by the duration of initial signaling, the number of agonist pMHC molecules, or the affinity of agonist molecules. Two general cases (defined in the methods) are studied: one in which the initial signal strength is large, and the other in which it is small; these values are defined more precisely in the context of each simulation.

Representative time courses are presented in Figs. 3 and 4. Consider first the behavior of calcium mobilization and its related transcriptional products (Figs. 3a,b). In the cases of low and high signal strengths, the activity of this pathway cycles approximately in phase with the cycling of the stimulus. This is because calcium mobilization and Erk activation are relatively fast in our model. For cases of weak stimulation, the signal cycles in phase with the duration of stimulation but is subject to large fluctuations (Fig. 3b) that may be interpreted as a less reliable signal.

In Figs. 3c,d, we focus our attention on the interaction of this pathway with the rest of the network—our results for the case where the stabilization of cFOS is cooperative are shown. In this case, the time courses for IEGs and cytokine production are very different from those showing Ca2+/NFAT activity. In Fig. 3c., IEGs slowly accumulate upon stimulation. Once the signal is disrupted, IEG accumulation halts and then resumes once the stimulus is reintroduced. Cytokine production (Fig. 3d.) then follows from the presence of IEGs; given a sufficient amount of IEG accumulation, cytokine is produced provided that the intermediates from the parallel pathway are active. On the contrary, for weak stimulation, there is, however, no IEG and cytokine production because the cooperative nature of the enzymatic reactions leads to the hyperphosphorylated stable form of cFos exhibiting an all or nothing response (data not shown). In the cooperative model, there is no response below a certain threshold of signal strength.

Now consider results from our computer simulations for the same model, but for the case where the stabilization of cFOS is not cooperative but rather occurs in a linear manner according to simple laws of mass action in the enzyme kinetics. Results from the simulations show that there is no qualitative difference in the cases of strong and weak signal. Only the relative amounts of chemical species produced are different in the two cases. In this case, we observe a memory effect in the computer simulation irrespective of the strength of the signal(data not shown).

Finally, we observe the case where IEG products are embedded in an autocatalytic feedback loop (Fig. 4). For strong stimulation, we see production of stable IEG products that prepares for cytokine production at a later time(Fig. 4a). However, when the stimulus is disrupted, the amount of IEG decays to a steady value during the period of interruption. When stimulation is reinitiated, the amount of cFOS continues to grow monotonically and its activity contributes to the immediate production of cytokine(Fig. 4b).

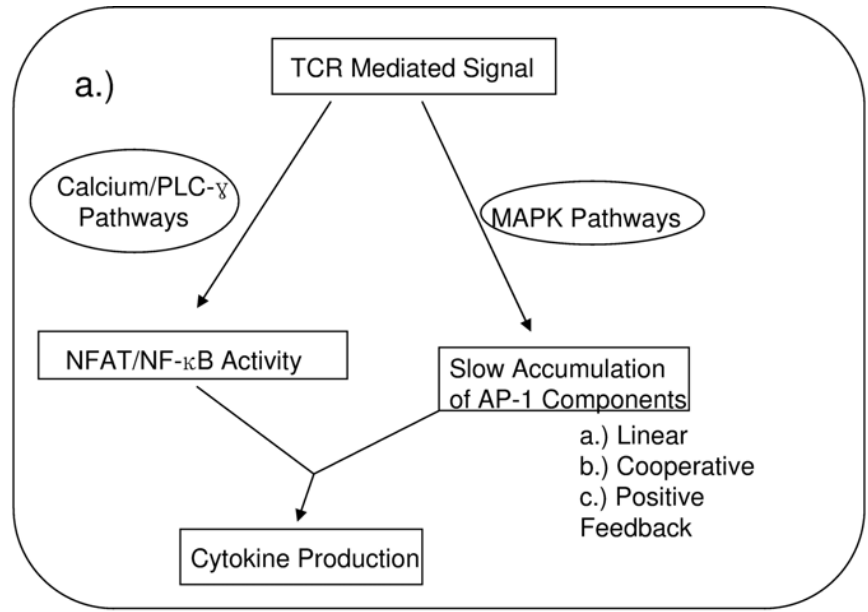

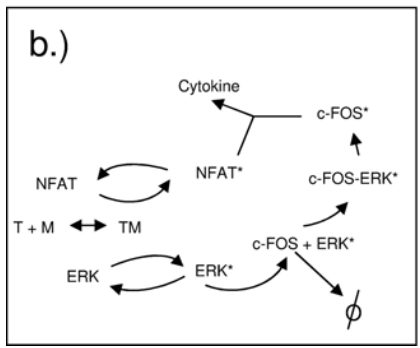 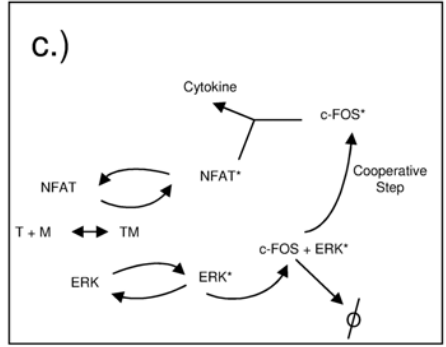 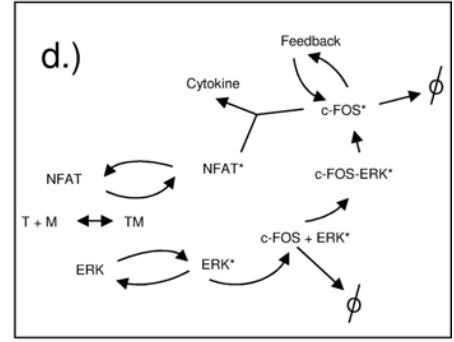

**Figure 2. Diagrams of the simplified signaling networks used in the computer simulations.** a.) An overall scheme for the signaling model to be simulated. Parallel pathways, whose activation occurs at different time scales, converge to produce cytokine. b–d.) Reaction schemes for each model, b.) linear c.) cooperative and d.) feedback induced models for persistent activity cFOS.
doi:10.1371/journal.pone.0000627.g002

**Table 1.** Parameters used in the Monte Carlo simulations

| Reaction | Rate Constant | Linear Model | | Feedback Model | |
| --- | --- | --- | --- | --- | --- |
| 1 | 0.0000001 | 10a | 0.1 | 10c | 0.1 |
| 2 | 0.019 | 11a | 0.1 | 11c | 0.1 |
| 3 | 0.01 | 12a | 0.1 | 12c | 0.1 |
| 4 | 0.1 | | | 13c | 0.1 |
| 5 | 0.01 | Cooperative Model | | $\alpha$ | 100 |
| 6 | 0.1 | $\alpha$ | 0.01 | $\beta$ | 600 |
| 7 | 0.0005 | $\beta$ | 600 | H | 10 |
| 8 | 0.1 | H | 10 | | |
| 9 | 0.000001 | | | | |

doi:10.1371/journal.pone.0000627.t001

Qualitative differences among the three models are further illustrated by monitoring the time evolution of probability distributions of pertinent signaling species. Such distributions are the analog to monitoring the statistics of the cell population. In Fig. 5, distributions of IEGs(Figs. 5a,b) and cytokines(Figs. 5c,d) produced at several time points are computed. Three time points are considered: at 30 minutes after the first round of signaling, at 50 minutes after the first period of interruption, and at 80 minutes after the second round of signaling.

In the presence of a feedback loop and sufficiently strong stimulation(Figs. 5a,c), we observe, at thirty minutes, a broadly peaked distribution centered on a large amount of IEGs (Fig. 5a). Little to no cytokine is produced at that time (Fig. 5c.). After signaling has been disrupted for 20 minutes, the simulated cell population of active IEGs shifts to the left and becomes sharply peaked. Now, at the end of the second round of signaling, the population remains sharply peaked and shifts markedly to the right and the number of IEGs and cytokines become greatly amplified(Figs. 5a,c). The feedback loop, in effect, allows for large signal amplification and reduces the amount of noise propagated in the signaling cascade(Figs. 5a,c).

For the case of weak stimulation(Figs. 5b,d), signal integration in the presence of a feedback loop shows very different qualitative behavior. After the first round of signaling, a broad distribution centered on a small amount of IEGs is observed (Fig. 5b). After the following twenty minutes of interrupted signaling, the entire population of IEGs decays to zero. The next round of signaling leaves the cell population identical to that which was observed at the end of the first round of signaling. Hence, the presence of a feedback loop along with rapid turnover of active signaling molecules suppresses all memory effects in the case of a sufficiently weak signal. These effects are also exhibited in the distributions of cytokine production(Figs. 5c,d). In the first case where cFOS exhibits memory of the previous exposure to stimulation (Fig. 5c), we observe that the amount of cytokine produced is highly amplified in the second round of signaling whereas after the first round, cytokine production is barely detectable. In the case where no memory effect is observed, only minute amounts of cytokine production are observed–the amount of active IEGs is insufficient to produce significant quantities of cytokine.

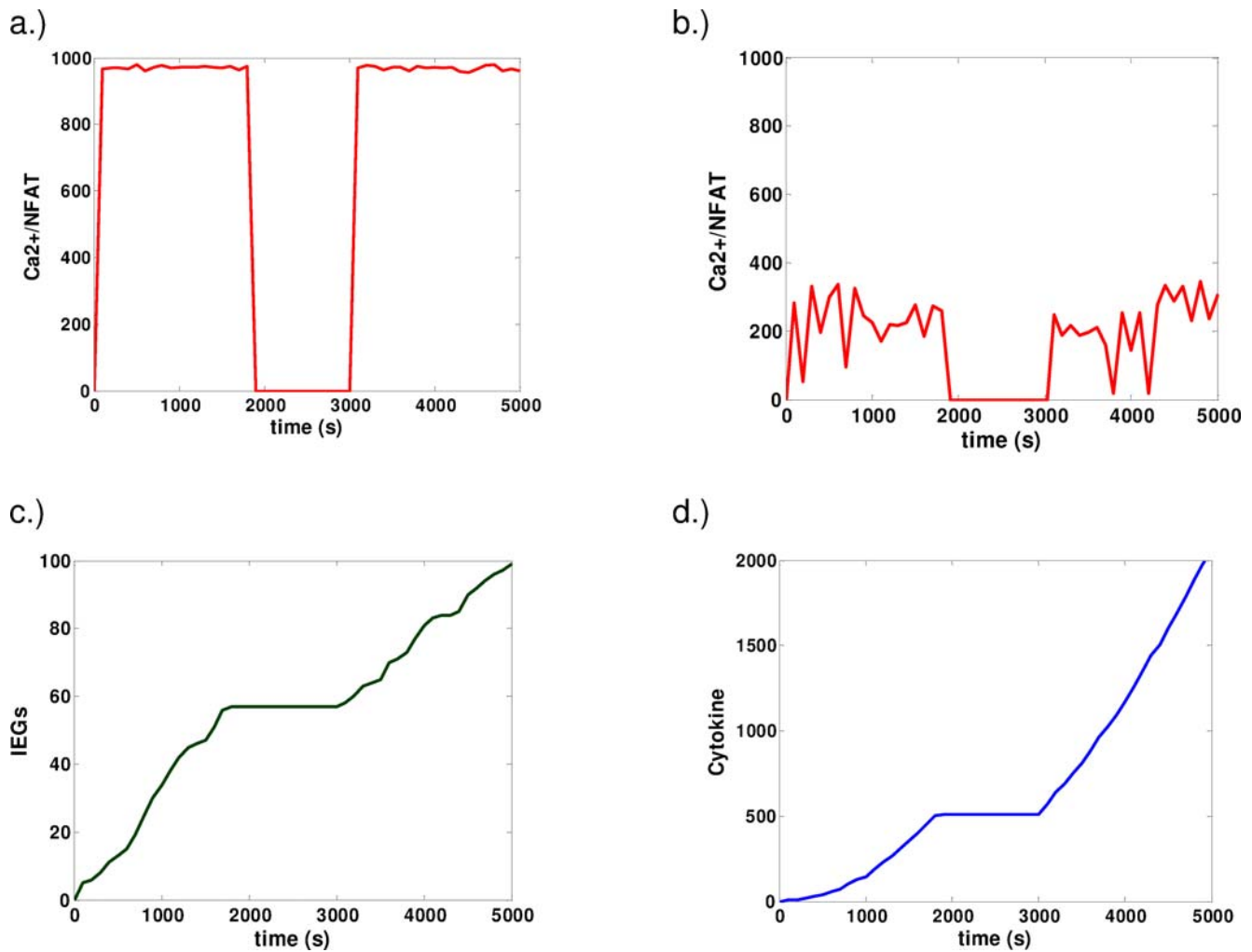

Figure 3. Representative dynamics for cooperative and linear models. a,b) Ca2+/NFAT dynamics. Under strong stimulation (a). Activity cycles roughly in phase with the duration of stimulation. Under weak stimulation (b), activity also cycles approximately in phase with the duration of signaling. However, such activity is less consistent than that observed in the case of strong stimulation and subject to large fluctuations. c,d.) Trajectories of active IEGs (e.g. cFOS) (c) and cytokine (d) for the case of cooperative cFOSp/Erkp dynamics in the presence of sufficiently strong stimulation. Other qualitatively similar cases are presented in the supporting online information.
doi:10.1371/journal.pone.0000627.g003

In the absence of a feedback loop (Figs. 5e,f), the distributions of active cFOS (Fig. 5e) at 50 minutes are maintained at the same levels as were obtained after the first 30 minutes of signaling. This is because IEGs simply remain stable in these cases. After the second round of signaling, the amount of active cFOS about doubles and cFOS activity is not highly amplified after the second round of signaling. The distribution of cytokine production (Fig. 5f) follows from this result. A small amount of cytokine is produced in the first round of signaling and is proceeded in the second round by a much larger amount–this follows from the activity of cFOS that was acquired in the first of signaling. Similar qualitative behavior is seen in all other cases (cooperative and linear models, data not shown) except when enzymatic reactions are cooperative and an insufficient amount of stimulation leads to no activity. In such a case, no active cFOS, cytokine, and hence no memory is obtained; such effects arise from the switch-like or "all or none" nature of the signaling circuit for the cooperative model.

These results appear to suggest that experiments that probe the dose-dependent response of the signaling system would give qualitatively different results for each of the three models. The presence of a feedback loop would allow for an "all or nothing" memory effect and cytokine response, as would the presence of a cooperative "switch-like" enzymatic stabilization mechanism.

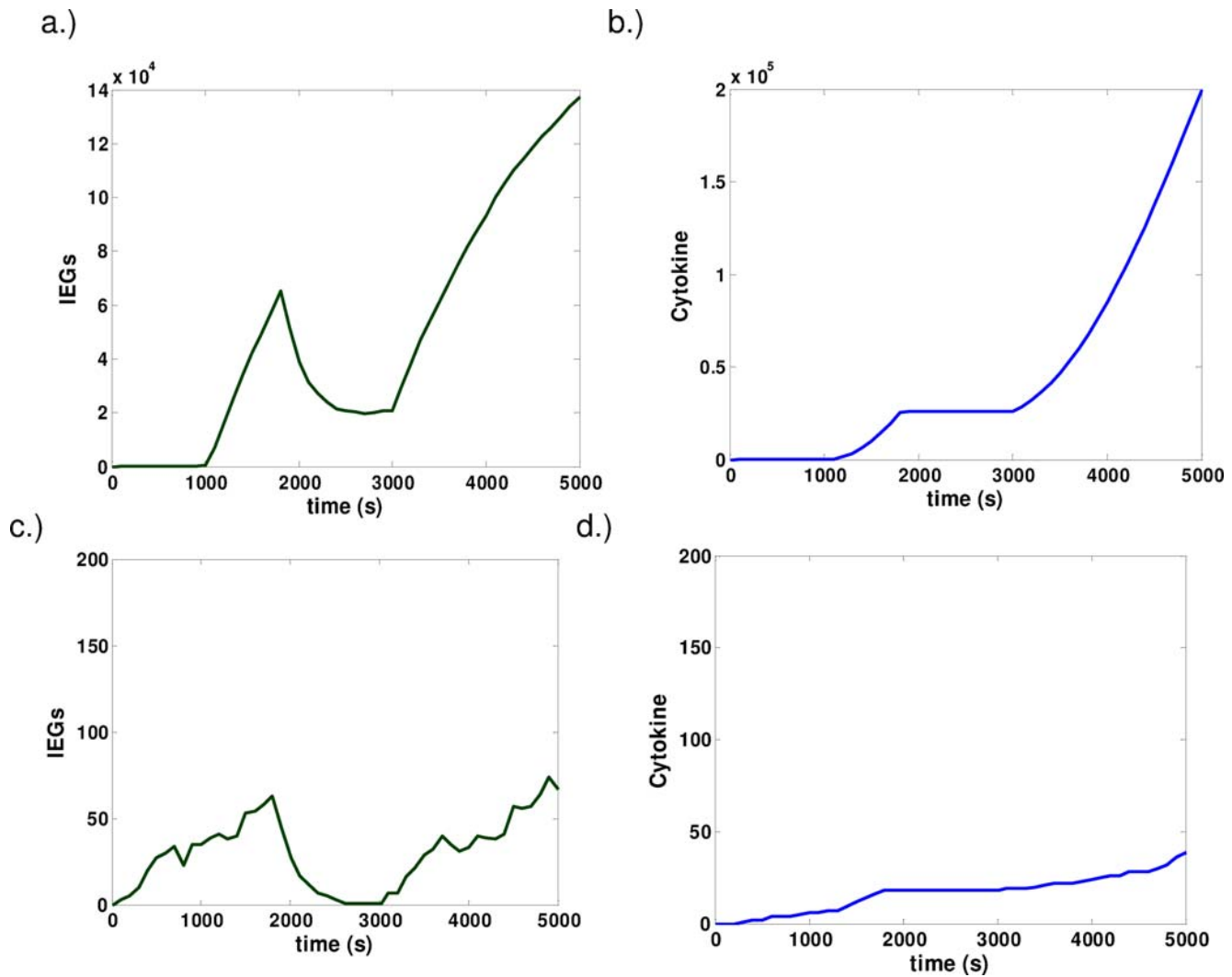

**Figure 4. Representative dynamics for models with positive feedback.** Scenarios involving strong (a,b) and weak (c,d) stimulation are considered. Representative timecourse for the activity of IEGs (a,c) and cytokine production (b,d) are shown. Memory effects are acquired in the case of strong stimulation. While weak stimulation triggers IEG activity, such activity is effectively reset upon interruption of the signal.
doi:10.1371/journal.pone.0000627.g004

Alternatively, in the case of a model with linear enzymatic kinetics, a graded response would be predicted. A graded dose response is expected because there is no mechanism to give rise to a threshold in the signaling network. Such behaviors are displayed in Fig. 6 that gives a comparison of the dose response curves. Note that in the case of the feedback loop (Fig. 6c), a hysteresis is observed–the behavior of the dose response curve going forward is different from that obtained going backwards ("going backwards" refers to starting initially in the state with a large amount of active IEG). This sensitivity of signal output to the initial physiological conditions provides the source of memory.

What might be the biological consequence of hysteretic effects present in the production of IEG products? We first consider the dependence of hysteresis on the strength of such a feedback loop. Such an effect in Fig. 6c implies that the persistence of memory effects in IEG accumulation can be made permanent. The backwards dose response curve in Fig. 6c indicates that, after twenty minutes of disrupted stimulus, such a memory effect for cytokine production will be apparent under all physiological conditions that may be realized during a subsequent round of signaling. In contrast, Fig. 7. considers the effects of decreasing the feedback strength on the hysteresis in the signaling circuit. As the strength of the feedback loop, i.e. the value of $\alpha$, decreases, the threshold signal strength required for acquisition of the memory effect increases and the curve markedly shifts to the right. Such a dependence of system behavior on the strength of the feedback could allow for some degree of plasticity in the response. For weaker feedback strengths, the dose response, while still retaining the switch-like characteristic, becomes reversible. Starting from the memory-competent state and decreasing signal strength, a point is reached at which the amount of active cFOS decays to zero (for $\alpha = 1, 2$ in Fig. 7). This implies that even if the first round of signaling is sufficient to induce such a memory with IEG products, a threshold amount of signal is required to achieve the memory effect. Therefore, cytokine production will only begin more quickly in subsequent rounds of signaling if the stimulation in that round is strong enough. It is interesting to speculate that such a control mechanism may serve to establish better specificity in the subsequent rounds of signaling.

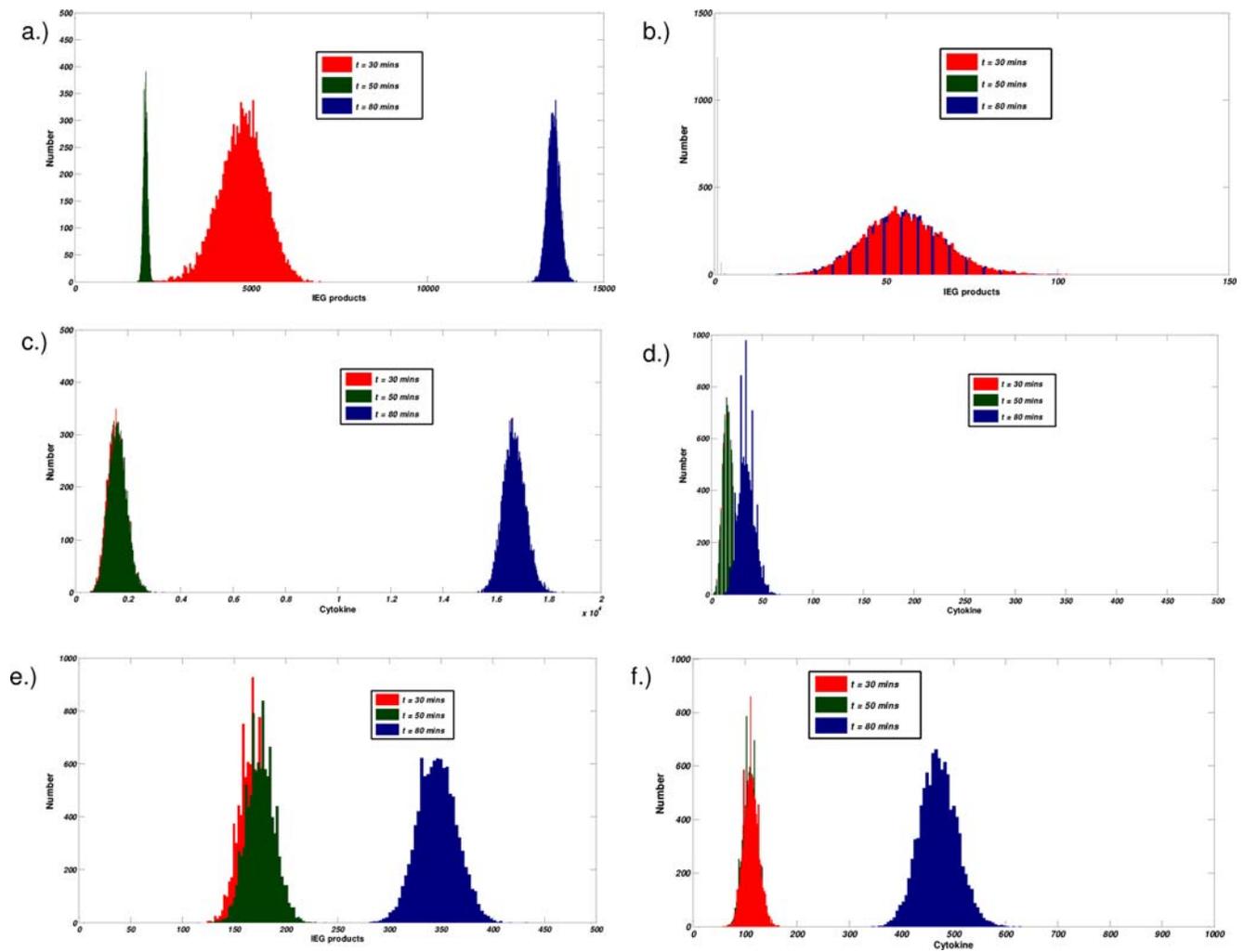

**Figure 5. Comparison of the distributions of active IEGs and Cytokine production for different models.** a–d, results from feedback model. e,f, results from a linear model. Probability distributions are computed at three time points, t = 30 minutes (after first round of stimulation) red , t = 50 minutes (after first period of interrupted signaling) green , and t = 80 minutes (after the completion of the second round of signaling) blue. IEG products (a,b,e) and Cytokine production (c,d,f) are considered. In the presence of a feedback loop, two separate cases (strong (a,c) and weak (b,d)) signal strength are analyzed.
doi:10.1371/journal.pone.0000627.g005

## DISCUSSION

Our computational analysis suggests specific experiments that could provide insights into the mechanisms that underlie the ability of T cells to integrate signals and retain a "memory" in the signaling process. The most significant experiments will be ones that monitor the stability of transcription factors in and out of the nucleus and determine whether individual activated molecules are stable or rather, constantly turning over when signal memory is exhibited. Signaling "memory" then can be assessed by the persistence of nuclear transcription factors after inhibition of the signaling pathway. Experiments with the Lck inhibitor PP2, in conjunction with immunofluorescence assays that make use of fluorescent secondary linked antibodies, can monitor the nuclear translocation of the relevant transcription factors such as Fos, Jun, NF-$\kappa$B, and NFAT upon disruption of TCR mediated signaling. These experiments will be essential to understanding the relevant transcription factors that enable short-term biochemical signaling memory.

Feedback loops are ubiquitous in T cell signaling[21] and evidence for bistability in signaling pathways has been shown in numerous cases[16]. Some of the most comprehensive studies involve studies of JNK signaling in Xenopus Oocytes[22]. These works demonstrate that the JNK pathway can both respond to stimuli in an all or none manner and exhibit all the features of a cascade involving strong positive feedback, including hysteresis. It is interesting to speculate that JNK signaling may exhibit similar features in T cells. The JNK cascade is involved in cytokine production, and exhibits many features of bistability. In CD8 T cells, the transcription factor c-JUN, a product of the JNK cascade, remains active for up to 24 hours after the stimulus has been removed[13]. However, JNK activity has recently been shown, in one case, to be unnecessary for IFN-$\gamma$ production. It is also possible that signaling leading to the production of protein products of IEGs (e.g., cFos) is embedded in a positive feedback loop.

Furthermore, one prediction obtained from this model suggests that one could distinguish between the two possible mechanisms for sustained activity by carrying out photobleaching experiments using GFP constructs of the relevant transcription factors. If signaling memory is due to a long half-life of the signaling intermediate, photobleaching will largely eradicate the ability to observe nuclear localization of the transcription. If, on the other hand, signaling memory is due to existence of positive feedback in the signaling circuit, then one would expect that nuclear fluorescence will rapidly recover after photobleaching.

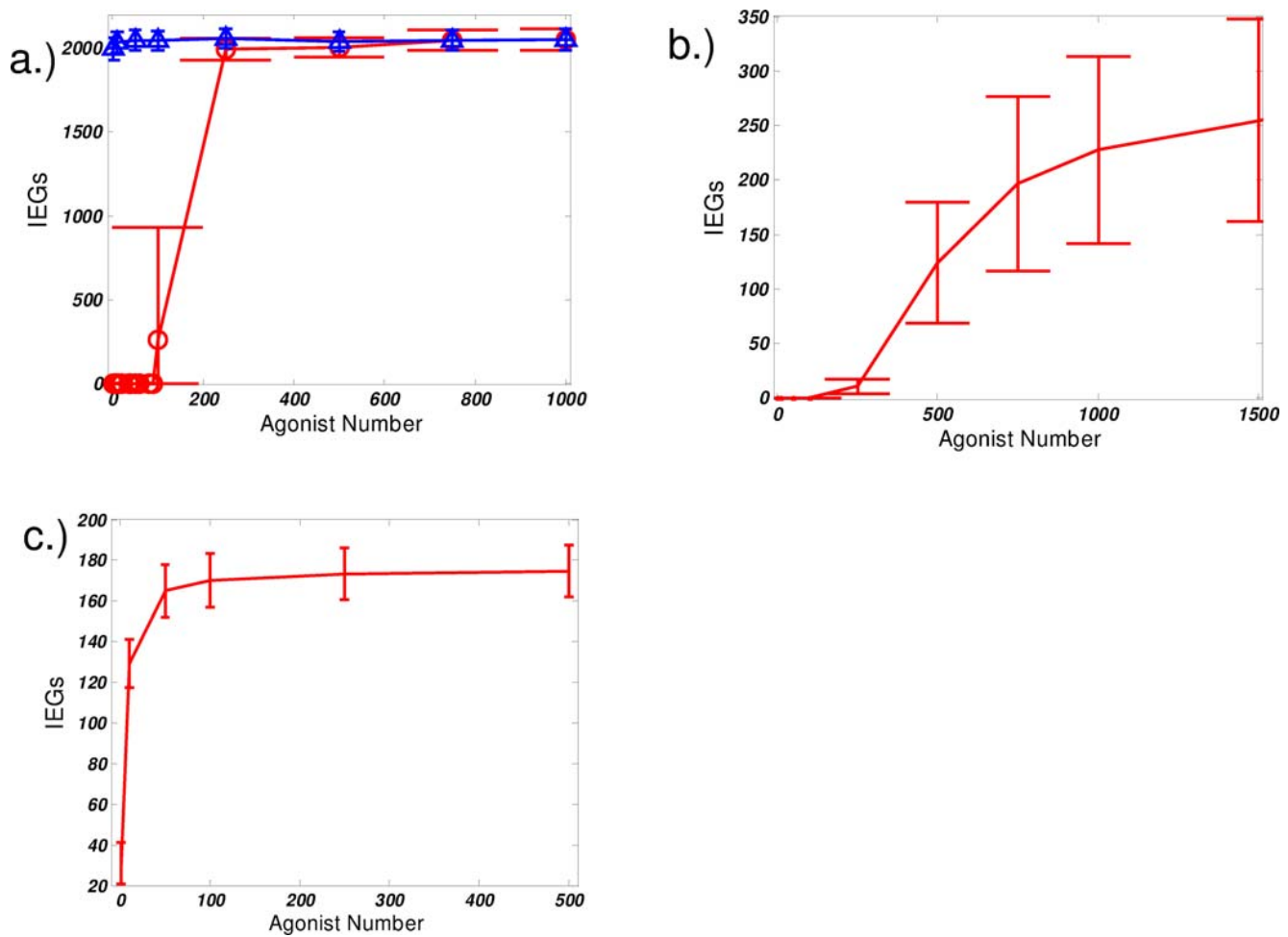

Figure 6. Comparison of dose response curves. Average values of active IEG products after the stimulus has been removed for 20 minutes (t = 50 minutes) a.) feedback regulation model. forward (red circles) and backward (blue triangles) curves are shown. Strong hysteresis is observed b.) cooperative reaction model c.) linear, mass-action kinetics reaction model "error bars" are computed by considering the standard deviation—one measure of the magnitude of noise in the signaling process.
doi:10.1371/journal.pone.0000627.g006

These experiments will first help to first determine whether AP1 is the transcription factor that is the source of biochemical memory, and should allow one to discriminate between the models of biochemical memory we have studied *in silico*. Should it be found that a particular model where individual activated molecules are rendered stable for long times is appropriate, the importance of cooperative enzymatic modifications can be determined by measurements of dose-response curves for the amount of activated cFos as a function of signal strength, which could be modulated by the amount or quality of the agonist pMHC as well as the duration of the initial signal.

Directly observable predictions about how the kinetics and strength of TCR signaling affects signaling memory could be tested. Since a model involving either a cooperative mechanism or a feedback loop predicts a threshold for a memory effect in signal transduction, the strength of signal will determine whether or not T cells can integrate signals during multiple exposures to antigen. These models propose that there exists some crossover between weak and strong agonists, short and long durations of TCR signaling, and concentrations of low and high numbers of agonist; this crossover will determine whether or not a lag time is required for cytokine production during subsequent rounds TCR signaling after the signal has been disrupted.

Signaling memory implies the persistence of sustained activity of some biophysical or molecular signaling intermediate even after signaling is interrupted by removing the stimulus. Such an intermediate could be the persistence of spatially clustered signaling components (e.g. as the formation of microclusters)[23,24] or the presence of compartmentalized signaling components[25] whose prior assembly constituted a rate limiting step. However, many of these processes are actin-mediated[23] and likely rapidly aborted in the absence of a signal. In light of these difficulties, it seems that a model invoking positive feedback is a plausible explanation for the molecular origins of memory in T-cell signal integration. Such a model is desirable on several bases; it provides noise reduction, plasticity in threshold tuning, precise control of signal amplitude and timing, and potentially useful hysteretic effects in the acquisition of such a signaling memory. The other models lack most, if not all, of these features. However, such memory effects in the form of spatial localization or perhaps time delays can not be excluded at this time.

In summary, we have explored, *in silico*, several molecular models which can explain the mechanism of biochemical memory in T cell signaling and activation. Each model involves the sustained activation of a certain transcription factor in the presence of disrupted signaling. Furthermore, our computer simulations make several predictions that we have briefly outlined.

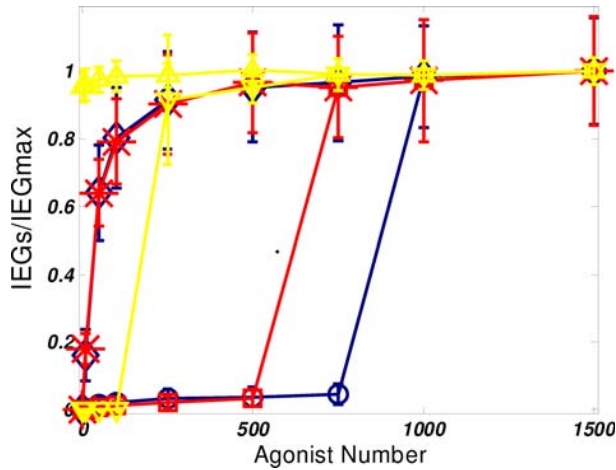

**Figure 7. Evaluation of the effects feedback strength.** Forward and backward dose response curves for varying feedback strengths, $\alpha=1$ (blue), $\alpha=2$ (red), and $\alpha=5$ (yellow). Different markers correspond to the forward and backward dose response. At high feedback strengths, the response is irreversible. At low feedback strengths, the active state can reverted back to the inactive state. Again, values are calculated at $t=50$ minutes.
doi:10.1371/journal.pone.0000627.g007

It is our hope that this work will serve to motivate as well as guide future experimentation into mechanisms underlying biochemical memory in T-cell signaling and activation. Once these mechanisms are better understood, further elaboration on the details of our computer models will be necessary to provide a better quantitative analysis of the mechanism governing the memory phenomenon. Also, it will be important to address in the near future how signaling memory at the cellular level functions in the context of T-cell activation in vivo where T-cell migratory patterns in lymph nodes are important in controlling the overall outcome of the physiological response. Integration of a more detailed computational model of the signaling pathways that maintain short-term memory with a computational model for T-cell trafficking in lymph nodes will be essential for understanding this problem. A model of this sort can then be used in conjunction with two-photon imaging experiments *in vivo* along with genetic and biochemical experiments to investigate the underlying mechanisms in T-cell activation across multiple length and time scales, from the molecular features governing the dynamics of signaling pathways to the clearance of infection occurring at higher levels of biological organization.

## METHODS

The signaling models that we chose to simulate consist of the following half reactions and the basic set of molecular processes common to each of the three models is as follows (cFOS is taken to be the example of the Immediate Early Gene product):

$$T+M \rightarrow TM \quad (1)$$

$$TM \rightarrow T+M \quad (2)$$

$$NFAT+TM \rightarrow NFAT^* + TM \quad (3)$$

$$NFAT^* \rightarrow NFAT \quad (4)$$

$$ERK+TM \rightarrow ERK^* + TM \quad (5)$$

$$ERK^* \rightarrow ERK \quad (6)$$

$$ERK^* + GENE \rightarrow cFOS + ERK^* \quad (7)$$

$$cFOS \rightarrow NULL \quad (8)$$

$$cFOS^* + NFAT^* \rightarrow Cytokine \quad (9)$$

Reactions determining cFOS* accumulation are shown for the linear (a), cooperative (b), and feedback (c) models.

a.) Linear model

$$cFOS + ERK^* \rightarrow cFOS - ERK^* \quad (10a)$$

$$cFOS - ERK^* \rightarrow cFOS + ERK^* \quad (11a)$$

$$cFOS - ERK^* \rightarrow cFOS^* + ERK^* \quad (12a)$$

b.) Cooperative model

$$cFOS + ERK^* \rightarrow cFOS^* + ERK^* \quad (10b)$$

c.) Feedback model

$$cFOS + ERK^* \rightarrow cFOS^* + ERK^* \quad (10c)$$

$$cFOS^* \rightarrow cFOS \quad (11c)$$

$$cFOS^* + GENE \rightarrow cFOS^* + FEEDBACK \quad (12c)$$

$$FEEDBACK \rightarrow NULL \quad (13c)$$

$$FEEDBACK + GENE \rightarrow cFOS^* + FEEDBACK \quad (14c)$$

The parameters used in the Monte Carlo simulations are in table 1.

We simulated these models by solving a master equation[26], $\frac{\partial P(\vec{n},t)}{\partial t} = \sum_{\vec{n}'} W_{\vec{n}' \vec{n}} P(\vec{n}',t) - \sum_{\vec{n}'} W_{\vec{n} \vec{n}'} P(\vec{n},t)$, whose solution gives the time evolution of the probability distribution for the system of chemical species to be in state $\vec{n}$. $\vec{n}$ is a vector whose components give the number of molecules of each molecular species. $W_{\vec{n} \vec{n}'}$ gives the transition probability per unit time for the transition of $\vec{n}$ to $\vec{n}'$. We used the standard stochastic simulation algorithm developed by Gillespie for solving master equations involving chemical reactions [27]. We constructed $W_{\vec{n} \vec{n}'}$ by first considering "mass action" kinetics that are determined by the topology of the reaction network corresponding to each signaling model. For the more complicated reaction mechanisms that we invoked to model cooperativity and feedback, we instead use the following unit-time transition probabilities, $W^{COOP}_{ncFOS^* \rightarrow (n+1)cFOS^*} = \frac{\alpha [cFOS][ERK^*]^H}{\beta^H + [ERK^*]^H}$ and

Biochemical Memory in T cells$$W^{FEED}_{ncFOS^* -> (n+1)cFOS^*} = \frac{\alpha[cFOS^*]^H}{\beta^H + [cFOS^*]^H}$$

respectively. $\alpha$ and $\beta$ are adjustable parameters that determine the strength of the nonlinear interaction. H determines the degree of cooperativity. Distributions were compiled from simulations of 10,000 statistically independent trajectories for each case presented. When plotting average behavior, error bars were obtained from simulations of 1000 trajectories. All code was written in ANSI C and compiled with the gnu C compiler, GCC.

The set of kinetic parameters used in the simulations is shown in Table 1. It is important to note that the simple signaling models we presented are not designed to quantitatively reproduce or fit experimental data; rather, their purpose is an attempt to lend deeper insight into the nature of such signaling mechanisms and generate useful predictions. However, our choice of parameters is not arbitrary; parameters were first estimated and constrained by way of a careful analysis of the important, experimentally measured time scales in the signaling process. Then, sensitivity of these parameters to the various mechanisms in question was studied.

## Sensitivity of model to changes in biophysical parameters

The mathematical models of cell signaling that we analyzed are comprised of several modular components. Therefore, the sensitivity of the qualitative results of our models to the choices of kinetic parameters may best be understood by considering the key competing time scales, $\{\tau_{sig}, \tau_{p1}, \tau_{p2}, \tau_{mem}, \tau_{cyt}\}$, that emerge in the modular network architecture that we constructed.

$\tau_{sig}$ is the time scale for signals derived from TCR-MHC to propagate to downstream messenger pathways. $\tau_{sig}$ emerges from kinetic constants and initial concentrations in reactions (1) and (2). $\tau_{sig}$ then, is a measure of the overall signal strength, which can be varied by adjusting the agonist concentration. For example, high strength (1000 pMHC molecules) and low signal strength (10 pMHC molecules) as well as long and short durations of signal map onto a value of $\tau_{sig}$. $\tau_{p1}$ and $\tau_{p2}$ are the characteristic time scales involved in activating the two parallel messenger pathways in our model. $\tau_{p1}$ is the time scale to activate the fast pathway (e.g. Ca2+ Mobilization and active NFAT). $\tau_{p2}$ is the time scale required to activate the other pathway that leads to the synthesis of unstable IEG products. $\tau_{mem}$ is the time needed to establish a biochemical memory in the signaling circuit. A model assumption is that $(\tau_{p1} \sim \tau_{p2}) \ll \tau_{mem}$. If this were not the case (i.e. $(\tau_{p1} \sim \tau_{p2}) > \tau_{mem}$) then subsequent rounds of signaling would not quickly produce cytokine. Thus, $\tau_{p1}$ and $\tau_{p2}$ as well as the the time scale for cytokine production $\tau_{cyt}$ then limits the speed at which productive signaling can recover from interrupted stimulation. A mechanism involving the stabilization of IEGs as a source of memory requires that $\tau_{mem}$ be large–at least on the order of minutes.

Parameters from each model contributing to $\tau_{mem}$ (i.e. those in reactions 10a–c, 11a,c, 12a–14c) were varied and results are either presented in the main text or are discussed below. For the linear model, $\tau_{mem}$ changes in response to the kinetic parameters in reactions 10a, 11a, 12a and subsequently determines the amount of stable IEG but does not affect qualitative findings. For the cooperative model, parameters in reaction 10b then control the positioning and sharpness of the threshold. $\alpha$ controls the amount of stable IEG that can be obtained. The behavior of the parameters involved in the feedback loop is discussed in the results section. Finally, $\tau_{cyt}$ is a time required for cytokine production once signaling intermediates (available IEGs and Transcription factors come that from the other pathway) are available. Changing $\tau_{cyt}$ then results in changing the amount of cytokine produced in the simulation in a monotonic fashion.

For the most part, we found that many of the qualitative results obtained from our models are robust to large (greater than 10-fold) variations in key individual parameters. However, two key parameters in our model could potentially change the qualitative results of our computer simulations; these parameters are the rate constant for the de novo synthesis of IEGs (reaction 7) and the rate constant for IEG decay (reaction 8). This is because, in our model, there exists a competition between the synthesis of cFOS and its decay. Upon varying these rates, we find that our key result can be sensitive to the rate of cFOS production. If cFOS synthesis is too slow, then no stabilized cFOS will be present once the stimulus is removed at t = 30 minutes in our simulations.

## ACKNOWLEDGMENTS

The author is especially grateful to Arup Chakraborty for his introduction to this problem as well as for his encouragement and helpful suggestions. Salvatore Valitutti, Shuyan Qi, and Andrey Shaw are also gratefully acknowledged for insightful comments on early versions of this work.## Author Contributions

Conceived and designed the experiments: JL. Performed the experiments: JL. Analyzed the data: JL. Contributed reagents/materials/analysis tools: JL. Wrote the paper: JL.## REFERENCES

1. Sumen C, Mempel SR, Mazo IB, von Andrian UH (2004) Intravital microscopy: Visualizing immunity in context. Immunity 21: 315–329.
2. Mempel TR, Henrickson SE, von Andrian UH (2004) T-cell priming by dendritic cells in lymph nodes occurs in three distinct phases. Nature 427: 154–159.
3. Faroudi M, Zaru R, Paulet P, Muller S, Valitutti S (2003) Cutting edge: T lymphocyte activation by repeated immunological synapse formation and intermittent signaling. Journal of Immunology 171: 1128–1132.
4. Munitic I, Ryan PE, Ashwell JD (2005) T cells in G(1) provide a memory-like response to secondary stimulation. Journal of Immunology 174: 4010–4018.
5. Lee KH, Dinner AR, Tu C, Campi G, Raychaudhuri S, et al. (2003) The immunological synapse balances T cell receptor signaling and degradation. Science 302: 1218–1222.
6. Altan-Bonnet G, Germain RN (2005) Modeling T cell antigen discrimination based on feedback control of digital ERK responses. Plos Biology 3: 1925–1938.
7. Li QJ, Dinner AR, Qi SY, Irvine DJ, Huppa JB, et al. (2004) CD4 enhances T cell sensitivity to antigen by coordinating Lck accumulation at the immunological synapse. Nature Immunology 5: 791–799.
8. Wylie DC, Das J, Chakraborty AK (2007) Sensitivity of T cells to antigen and antagonism emerges from differential regulation of the same molecular signaling module. Proceedings of the National Academy of Sciences of the United States of America 104: 5533–5538.
9. Dong C, Davis RJ, Flavell RA (2002) MAP kinases in the immune response. Annual Review of Immunology 20: 55–72.
10. Goldsmith MA, Weiss A (1988) Early Signal Transduction by the Antigen Receptor without Commitment to T-Cell Activation. Science 240: 1029–1031.
11. Murphy LO, Smith S, Chen RH, Fingar DC, Blenis J (2002) Molecular interpretation of ERK signal duration by immediate early gene products. Nature Cell Biology 4: 556–564.
12. Schade AE, Levine AD (2004) Cutting edge: Extracellular signal-regulated kinases 1/2 function as integrators of TCR signal strength. Journal of Immunology 172: 5828–5832.
13. Rosette C, Werlen G, Daniels MA, Holman PO, Alam SM, et al. (2001) The impact of duration versus extent of TCR occupancy on T cell activation: A revision of the kinetic proofreading model. Immunity 15: 59–70.
14. Arendt CW, Albrecht B, Soos TJ, Littman DR (2002) Protein kinase C-theta: signaling from the center of the T-cell synapse. Current Opinion in Immunology 14: 323–330.
15. Winslow MM, Neilson JR, Crabtree GR (2003) Calcium signalling in lymphocytes. Current Opinion in Immunology 15: 299–307.PLoS ONE | www.plosone.org    10    July 2007 | Issue 7 | e627

16. Ferrell JE (2002) Self-perpetuating states in signal transduction: positive feedback, double-negative feedback and bistability. Current Opinion in Cell Biology 14: 140–148.
17. Janeway C (2004) Immunobiology: Garland Science.
18. Sasagawa S, Ozaki Y, Fujita K, Kuroda S (2005) Prediction and validation of the distinct dynamics of transient and sustained ERK activation. Nature Cell Biology 7: 365–U331.
19. MacKeigan JP, Murphy LO, Dimitri CA, Blenis J (2005) Graded mitogen-activated protein kinase activity precedes switch-like c-fos induction in mammalian cells. Molecular and Cellular Biology 25: 4676–4682.
20. Gao M, Labuda T, Xia Y, Gallagher E, Fang D, et al. (2004) Jun turnover is controlled through JNK-dependent phosphorylation of the E3 ligase itch. Science 306: 271–275.
21. Reth M, Brummer T (2004) Feedback regulation of lymphocyte signalling. Nature Reviews Immunology 4: 269–277.
22. Xiong W, Ferrell JE (2003) A positive-feedback-based bistable 'memory module' that governs a cell fate decision. Nature 426: 460–465.
23. Yokosuka T, Sakata-Sogawa K, Kobayashi W, Hiroshima M, Hashimoto-Tane A, et al. (2005) Newly generated T cell receptor microclusters initiate and sustain T cell activation by recruitment of Zap70 and SLP-76. Nature Immunology 6: 1253–1262.
24. Grakoui A, Bromley SK, Sumen C, Davis MM, Shaw AS, et al. (1999) The immunological synapse: A molecular machine controlling T cell activation. Science 285: 221–227.
25. Daniels MA, Teixeiro E, Gill J, Hausmann B, Roubaty D, et al. (2006) Thymic selection threshold defined by compartmentalization of Ras/MAPK signalling. Nature 444: 724–729.
26. Van Kampen NG (2001) Stochastic Processes in Physics and Chemistry: Elsevier Science Publishing.
27. Gillespie DT (1976) General Method for Numerically Simulating Stochastic Time Evolution of Coupled Chemical-Reactions. Journal of Computational Physics 22: 403–434.